%% file: neurips_2024.tex
\documentclass{article}

\PassOptionsToPackage{numbers, sort&compress}{natbib}

\usepackage[preprint]{neurips_data_2024}

\usepackage[utf8]{inputenc} 
\usepackage[T1]{fontenc}    
\usepackage{hyperref}       
\usepackage{url}            
\usepackage{booktabs}       
\usepackage{amsfonts}       
\usepackage{nicefrac}       
\usepackage{microtype}      
\usepackage{xcolor}         
\usepackage{graphicx}
\usepackage{amsmath}
\usepackage{longtable}
\usepackage{array}
\usepackage{lscape}

\title{Benchmarking Out-of-Distribution Generalization Capabilities of DNN-based Encoding Models for the Ventral Visual Cortex.}

\author{{\large \bf Spandan Madan} \\
  Harvard University
  \And {\large \bf Will Xiao} \\
  Harvard Medical School
  \And {\large \bf Mingran Cao}\\
  Francis Crick Institute
  \AND {\large \bf Hanspeter Pfister} \\
  Harvard University
  \And {\large \bf Margaret Livingstone}\\
  Harvard Medical School
  \And {\large \bf Gabriel Kreiman} \\
  Harvard Medical School\\}

\begin{document}

\maketitle

\begin{abstract}
  We characterized the generalization capabilities of DNN-based encoding models when predicting neuronal responses from the visual cortex. We collected \textit{MacaqueITBench}, a large-scale dataset of neural population responses from the macaque inferior temporal (IT) cortex to over $300,000$ images, comprising $8,233$ unique natural images presented to seven monkeys over $109$ sessions. Using \textit{MacaqueITBench}, we investigated the impact of distribution shifts on models predicting neural activity by dividing the images into Out-Of-Distribution (OOD) train and test splits. The OOD splits included several different image-computable types including image contrast, hue, intensity, temperature, and saturation. Compared to the performance on in-distribution test images---the conventional way these models have been evaluated---models performed worse at predicting neuronal responses to out-of-distribution images, retaining as little as $20\%$ of the performance on in-distribution test images. The generalization performance under OOD shifts can be well accounted by a simple image similarity metric---the cosine distance between image representations extracted from a pre-trained object recognition model is a strong predictor of neural predictivity under different distribution shifts. 
  The dataset of images, neuronal firing rate recordings, and computational benchmarks are hosted publicly at: \href{https://drive.google.com/drive/folders/1OZQdPY6km6alH20mu5E6X_9Ke6HnHQAg?usp=share_link}{MacaqueITBench Link}. 
\end{abstract}

\input{introduction}

\input{related_works}

\input{dataset}

\input{constructing_ood_splits}

\input{quantifying_distribution_shifts}

\input{model_training_and_evaluation}

\input{results}

\input{conclusions}

\input{limitations}

\section{Acknowledgments}
This research was partially supported by NSF grant IIS-1901030. We thank Pranav Misra, Fenil Doshi and Elisa Pavarino for insightful discussions, and Harshika Bisht for design feedback on the figures. M.S.L. took the photos in the image set collected in the lab.

\bibliographystyle{unsrt}
\bibliography{references}

\newpage

\begin{center}
    \LARGE \bfseries Supplementary Material
\end{center}

\input{supplement}
\end{document}

%% file: introduction.tex
\section{Introduction}\label{sec:introduction}
\begin{figure}[t]
\centering
        \includegraphics[width=\textwidth]{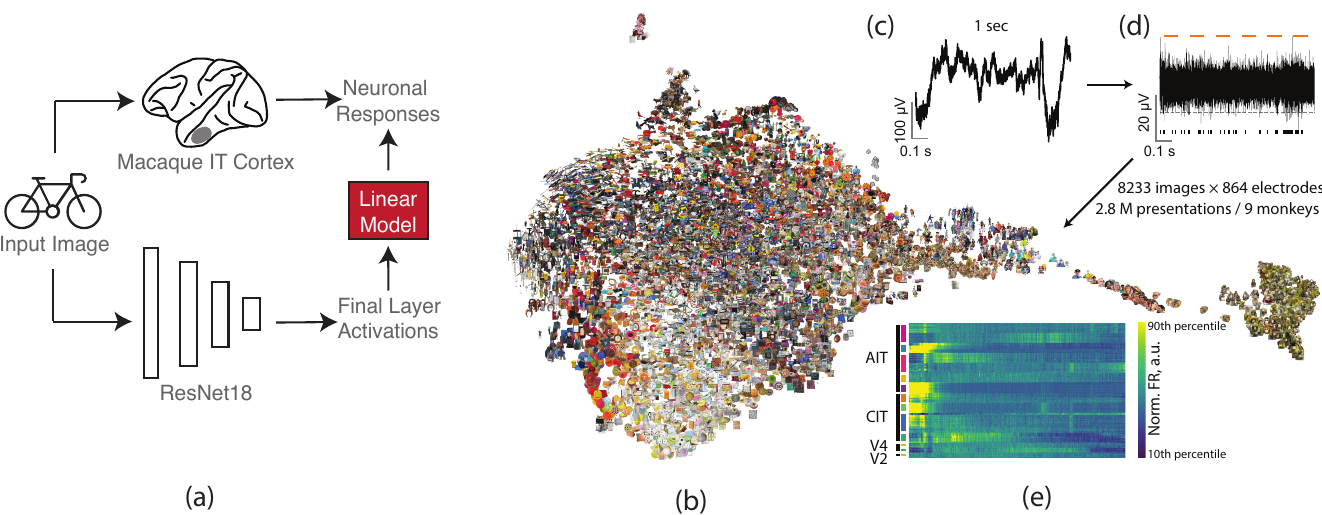}
        \caption{\textit{Modeling the visual cortex with MacaqueITBench}. (a) DNN-Based models of the visual cortex employ a linear model to map image features extracted from pre-trained DNNs (e.g., ResNet18) to neuronal responses collected from the macaque IT cortex. (b) A UMAP~\cite{mcinnes2018umap} visualization of the representation by the neural pseudo-population. Nearby images have more similar population responses. (c) An example one-second segment of the raw wideband signals recorded on an electrode. (d), The wideband signals were highpass filtered, and threshold-crossing events below a voltage value (horizontal dashed line) were counted as multiunit spikes (lower vertical ticks). The top horizontal bars indicate image presentation periods. (e) The heatmap shows the neural response matrix. Each row indicates the responses from an electrode, pooled across sessions. The columns correspond to images, sorted by the reverse UMAP horizontal order. The vertical bars to the left of the heatmap denote the recorded areas (black lines) and monkeys (colored lines).}
        \label{fig:fig_schematic}
\end{figure}

Deep Neural Networks (DNNs) for vision have internal representations that share similarities with neural representations in the visual cortex, including the primate ventral visual stream~\cite{bashivan2019neural,ponce2019evolving}. This representational similarity allows for models that use image representations extracted from a pre-trained DNN (e.g., ResNet~\cite{he2016deep}) to predict neuronal firing rates \cite{yamins2014performance} (Fig.~\ref{fig:fig_schematic}(a)).
However, DNNs are known to struggle with generalization under distribution shifts such as Out-of-Distribution (OOD) viewpoints~\cite{madan2023adversarial, madan2022and, cooper2021out}, materials and lighting ~\cite{madan2024improving, sakai2022three}, and noise~\cite{hendrycks2019benchmarking, croce2023seasoning}. 
This difficulty in generalization may also affect models of the visual cortex that rely on a DNN to extract image representations. 

We posit that, even within an image set where DNN-based models predict neural responses well under random splits across images, specific train-test splits with distribution shifts will impair model performance, proportional to the size of distribution shift. To test this hypothesis, we collected \textit{MacaqueITBench}, a large-scale dataset of responses to natural images by neurons in the macaque ventral visual pathway. The dataset represents neurons in V2, V4, Central IT (CIT), and Anterior IT (AIT) (primarily CIT and AIT) and responses to over $300,000$ images ($8,233$ unique images presented to seven monkeys over $109$ sessions), as illustrated in Fig.~\ref{fig:fig_schematic}(b).

Using \textit{MacaqueITBench}, we investigated the impact of distribution shifts on the neural predictivity of DNN-based models of the visual cortex. We constructed various OOD distribution shifts, some of which are schematized in Fig.~\ref{fig:constructing_splits}. 
Foreshadowing, our main finding is that distribution shifts in even low-level image attributes break DNN-based models of the visual cortex. This observation highlights a problem in modern models of the visual cortex---good predictions are limited to images that belong to the training data distribution. 

To explain the OOD model-performance drop, we built on theoretical work positing that generalization performance is closely correlated with the amount of distribution shift~\cite{canatar2021out, patil2024optimal}. 
While theoretical studies have examined simplistic, simulated data, we show that a suitable metric of the size of distribution shifts can account for the OOD generalization performance of neural-encoding models.

In summary, our main contributions are threefold:
\begin{itemize}
    \item We present \textit{MacaqueITBench}, a large-scale dataset of neural population responses to over $300,000$ images spanning multiple areas of the primate ventral visual pathway.
    \item We show that modern models of the visual cortex do not generalize well---simple distribution shifts can reduce neural predictivity to as low as $20\%$ of in-distribution performance.
    \item We show that a simple metric of distribution shift sizes can predict OOD neural predictivity.
\end{itemize}

%% file: related_works.tex
\section{Related Work}\label{sec:related_work}
\subsection{DNN-based models of the Visual Cortex}
A touchstone for visual neuroscience is the ability to predict neuronal responses to arbitrary images. On this test, DNN-based models have emerged as state-of-the-art models, best explaining neural responese across species---e.g., mouse and macaque---and visual cortical areas---from the primary visual cortex (V1) to the high-level inferior temporal cortex (IT).  DNN encoding models of the visual cotext are reviewed more generally in \cite{kriegeskorte2015deep,yamins2016using}.  Most pertinently here, these DNN-based models have been evaluated using random cross-validation (e.g., \cite{schrimpf2018brain}), which tests IID generalization. OOD generalization in such models have been sparsely examined; we are only aware of one study \cite{ren2023well}  comparing model fit to neural responses on two image types. Here, we systematically vary the type and degree of OOD splits to investigate how different splits lead to different generalization gaps.

\subsection{Out-of-distribution generalization capabilities of DNNs}
DNNs for object recognition have been documented to fail at generalizing across a wide range of distribution shifts. Such shifts include 
2D rotations and shifts~\cite{zhang2019making, chaman2021truly}, commonly occurring 
blur or noise patterns~\cite{hendrycks2019benchmarking, mintun2021interaction, xie2020adversarial, xie2020self}, and real-world changes in scene lighting~\cite{madan2021small,beery2018recognition, zhang2021adversarial}, viewpoints~\cite{madan2022and, barbu2019objectnet, liu2018beyond, zeng2019adversarial, madan2021small,cooper2021out, sakai2021three}, geometric modifications~\cite{belder2022random, xiao2019meshadv, yang2018realistic}, color changes~\cite{joshi2019semantic,shamsabadi2020colorfool}, and scene context~\cite{bomatter2021pigs,zhang2020putting}.

There have been three broad approaches to address the lack of OOD generalization in DNNs: first, modifying the learning paradigm including modifying the architecture or loss function to enforce invariant representations~\cite{arjovsky2019invariant, erfani2016robust, chattopadhyay2020learning, wang2021respecting, li2017deeper}, or using ensemble and meta-learning~\cite{li2018learning, balaji2018metareg, zhou2021domain}; second,  modifying the training data using data augmentation~\cite{zhang2017mixup, hendrycks2019augmix, xu2020randconv, huang2017adain}, or by increasing data diversity~\cite{xie2020adversarial, shankar2018generalizing, qiao2020learning, sinha2017certifying, NEURIPS2018_1d94108e}; third, scaling data up to beyond billions of data points~\cite{billion_scale,radford2021learning,oquab2023dinov2}. Despite these efforts, OOD generalization remains an unsolved problem for deep networks.

%% file: dataset.tex
\section{MacaqueITBench: Image-response recordings from the ventral stream}\label{sec:dataset}

We collected a large-scale dataset of neural population responses to over $300,000$ images across sessions, comprising $8,233$ unique natural images presented to seven monkeys over $109$ sessions. In each session, a monkey maintained fixation while images were rapidly presented in random order. Each presentation was $83$ milliseconds; with $83$--$150$ milliseconds between presentations.

The images derived from published image sets \cite{konkle2010conceptual} and photos taken in the lab and contained pictures of common objects, people, and other animals including monkeys (Fig.~\ref{fig:fig_schematic}(b)). Image thumbnails are shown in Fig.~\ref{fig:fig_schematic}(b)); sample images are provided in the supplement. Images belonged to over $300$ semantic categories annotated by hand. A full list of categories can be found in the supplement. The large number and diversity of images allowed us to construct various OOD splits. 

Neural responses were recorded on intracranial microelectrodes measuring extracellular electrical potentials (Fig.~\ref{fig:fig_schematic}(c)) pre-processed to extract multi-unit spiking activity (Fig.~\ref{fig:fig_schematic}(d)) \cite{buzsaki2012origin,super2005chronic}. The analyses included 640 electrodes (12 multi-electrode arrays) recorded in nine hemispheres of seven monkeys, spanning four ventral-stream areas: V2, V4, central IT (CIT), and anterior IT (AIT), primarily sampling CIT and AIT (Fig.~\ref{fig:fig_schematic}(e)).  The electrodes were chronically implanted, and the responses showed stable selectivity when pooled across sessions. Nevertheless, our modeling focused on the more finely resolved within-session trial-averaged responses. 


        

%% file: constructing_ood_splits.tex
\section{Constructing out-of-distribution data splits}\label{sec:constructing_ood}
\begin{figure}[t]
\centering
        \includegraphics[width=\textwidth]{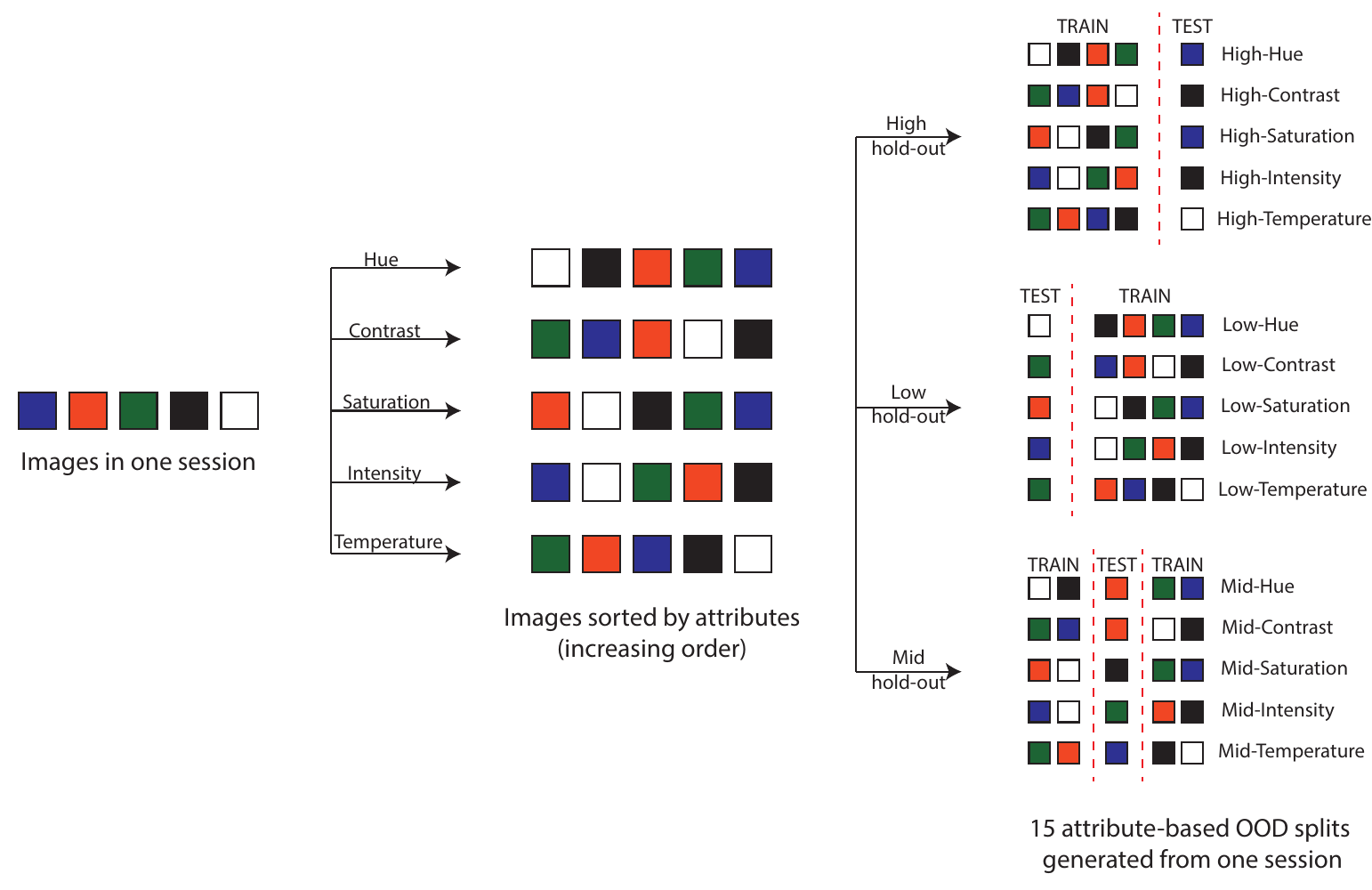}
        \caption{\textit{Constructing multiple attribute-based OOD splits}. For each of our $109$ sessions, we construct $15$ different attribute-based OOD splits. These correspond to $3$ hold-out strategies (\textit{high, low, mid}) for each of $5$ image-computable attributes (hue, contrast, saturation, intensity, temperature). For each attribute (e.g., hue), we compute the attribute value for each image in the session. For the \textit{high} hold-out strategy, all images with the attribute value above a percentile cut-off serve as the OOD test set with the remaining serving as the train set. Analogously for the \textit{low} hold-out splits, images below a percentile cut-off serve as the test set with the remaining serving as the train set. For \textit{mid} hold-out splits, images within the middle percentiles serve as the test set.
        }
        \label{fig:constructing_splits}
\end{figure}

We build on past work studying generalization under systematic distribution shifts~\cite{madan2022and, madan2024improving, arjovsky2019invariant, hendrycks2019benchmarking}, and define the training and test distributions parametrically using image attributes. Using these parametric data distributions, we construct three kinds of train-test splits: 


\textbf{InDistribution (InD) splits}: For each session, we created one In-Distribution (InD) split to compare with OOD generalization performance. We sampled $25\%$ of the images at random, and held these out as the InD test set, with the remaining serving as the training set.

\textbf{Attribute-based OOD splits:} For concreteness, we describe OOD splits based on image contrast; splits based on the other image attributes were constructed analogously. For each session, we computed the contrast value for each image. Then, one of three strategies were employed:

\begin{itemize}
    \item \textit{High hold-out}: The $75^{th}$ percentile of contrast values served as the cut-off. Images with contrast above the cut-off formed the test set. Remaining images formed the training set.
    \item \textit{Low hold-out}: The $25^{th}$ percentile served as the cut-off. All images below this served as the held-out test set. The remaining served as the training set. 
    \item \textit{Mid hold-out}: Images with contrast values between the $42.5^{th}$ and $62.5^{th}$ percentile served as the held-out test set. The remaining formed the training set.
\end{itemize}



\textbf{Cosine Distance-based splits:} To investigate the relationship between the size of distribution shift and neural predictivity, we constructed $3$ additional test splits. We first extracted the features for every image from the pre-final layer of a pre-trained ResNet18. A random image was picked to be the seed, and all images in the session were sorted in order of increasing cosine distance between the ResNet extracted features of the images and the seed. The sorted images were then divided into three chunks based on percentile cut-offs. The first chunk corresponds to the bottom $80$ percentile which served as the Training + In-Distribution Test split. A random subset of this first chunk was held out to form the In-Distribution test split, with the remaining serving as the training set. The second chunk is images in the $90^{th}$ to $95^{th}$ percentile, which are held-out as the \textit{Near-OOD} test split. Finally, the third chunk corresponds to images above the $95^{th}$ percentile. These are held-out as the \textit{Far-OOD} split. To ensure a gap between the train and test distributions, we discard images between the $80^{th}$ and the $90^{th}$ percentile. Note that the number of images in the In-Distribution test split was kept the same number of images as the Near-OOD split.



%% file: quantifying_distribution_shifts.tex
\section{Quantifying distribution shifts}\label{sec:quantifying_ood}
We present a unified framework for measuring distribution shifts over the parametric OOD train-test splits presented in Sec.~\ref{sec:constructing_ood}. 

\subsection{Representations for training and testing data-splits}

Let $D_T=\{i_1^T, i_2^T,...,i_N^T\}$ denote a train split of $N$ images, and let $D_t=\{i_1^t, i_2^t,...,i_n^t\}$ denote the corresponding test split of $n$ images. $\mathcal{R(.)}$ is a representation function that provides a vector representation for an image. The train and test images thus correspond to $\mathcal{R}(D_T) = \{\mathcal{R}(i_1^T), \mathcal{R}(i_2^T), \ldots, \mathcal{R}(i_N^T)\}$ and $\mathcal{R}(D_t) = \{\mathcal{R}(i_1^t), \mathcal{R}(i_2^t), \ldots, \mathcal{R}(i_n^t)\}$.

We analyzed representations $\mathcal{R}(i_j)$ formed by the features extracted for an image $i_j$ by a pre-trained DNN. We explore $8$ different DNN architectures, and multiple layers for every architecture. Equations below are agnostic to the architecture and the layer used. Other alternatives could include using HOG~\cite{dalal2005histograms} or GIST~\cite{oliva2006building} image features, or the vectorized pixel values of the image.

\subsection{Defining distances over different datasets}

To compute the shift between $\mathcal{R}(D_T)$ and $\mathcal{R}(D_t)$, we compared three distance metrics:

\textbf{Maximum Mean Discrepancy ($D_{MMD}$):} The MMD between the two datasets can be computed as

\begin{align*}
{D^2_{MMD}}(D_T, D_t) = \frac{1}{N^2} \sum_{j=1}^{N} \sum_{k=1}^{N} K(\mathcal{R}(i_j^T), \mathcal{R}(i_k^T)) + \frac{1}{n^2} \sum_{j=1}^{n} \sum_{k=1}^{n} K(\mathcal{R}(i_j^t), \mathcal{R}(i_k^t)) \\ - \frac{2}{Nn} \sum_{j=1}^{N} \sum_{k=1}^{n} K(\mathcal{R}(i_j^T), \mathcal{R}(i_k^t))
\end{align*}

Here, $K(\mathcal{R}(i_j^T), \mathcal{R}(i_k^t))$ is a kernel distance between the representations of images $i_j^T$ and $i_k^t$. A common choice for the kernel function $K(\cdot,\cdot)$ is the Gaussian RBF.

\textbf{Covariate-Shift ($D_{Cov}$):} 
Let $P_T(X)$ and $P_t(X)$ denote the distributions of the train and test input variables (i.e., image representations), and let $P(Y|X)$ denote the conditional distribution of the output (i.e., neural responses) given the input. Covariate shift exists if $P_T(X) \neq P_t(X)$ but $P_T(Y|X) = P_t(Y|X)$. $D_{Cov}$ can be computed by training a binary classifier to classify if data comes from the training or the testing dataset. We denote the accuracy of this classifier as $a_{T,t}$ and measure the covariate shift as:

\begin{align*}
 {D_{Cov}}(D_T, D_t) = 2 \times (0.5 - a_{T,t})).
\end{align*}




\textbf{Closest Cosine Distance ($D_{CCD}$):} For every image in the test set, we find its distance to the closest training image, and compute the mean of this distance over all test images. For brevity, we will refer to this as \textit{Closest Cosine Distance}. Let $i_k^T \in D_T$ denote the closest training image to test image $i_j^t \in D_t$ as measured by the cosine distance $D_{\text{cos}}(\mathcal{R}(i_j^t), \mathcal{R}(i_k^T))$. The distance $D_{\text{cos}}$ between two vectors $u$ and $v$ is given by

\begin{align*}
    D_{\text{cos}}(u, v) = 1 - \frac{u \cdot v}{\|u\| \|v\|}
\end{align*}

The average distance to the closest training image is

\begin{align*}
    D_{CCD} = \frac{1}{n} \sum_{j=1}^{n} \min_{k \in \{1, 2, \ldots, N\}} D_{\text{cos}}(\mathcal{R}(i_j^t), \mathcal{R}(i_k^T))
\end{align*}

%% file: model_training_and_evaluation.tex
\section{Model training and evaluation}\label{sec:model_training}
As depicted in Fig.~\ref{fig:fig_schematic}(a), we employ a linear model to map pre-trained model activations to neuronal firing rates from the IT cortex (Fig.  ~\ref{fig:fig_schematic}(a)). The linear model was learned using ridge regression. We used only pre-trained DNNs, not DNNs fine-tuned for our analysis.

For feature extraction, we investigated $8$ DNN architectures and $2$ layers for each architecture. The DNNs include supervised models trained on ImageNet (ResNet-18~\cite{he2016deep}, ViT~\cite{dosovitskiy2020image}), self-supervised models trained on billion-scale data with self-supervised and weakly supervised learning (ResNet18\_swsl~\cite{billion_scale}, ResNext101\_32x16d\_swsl~\cite{billion_scale}, ResNet-50\_ssl~\cite{billion_scale}), Noisy student with EfficientNet~\cite{xie2020selftraining}, self-supervised learning over billions of tokens (DinoV2~\cite{oquab2023dinov2}), and the multi-modal vision-language model CLIP~\cite{radford2021learning}.

A linear encoding model was fit for the trial-averaged responses of each neuron in a session. The results are presented as the mean and S.E.M. across $109$ sessions ($7$ monkeys); each session's results is the median across neurons. The model fit per neuron was quantified as the ceiling-normalized, squared Pearson's correlation, $r^2_\text{pred}/r^2_\text{cons}$ following convention~\cite{schrimpf2018brain,xiao2024feature} and related to the explained variance, $R^2$. The ceiling $r_\text{cons}$ of a neuron was calculated as its response correlation between split-half trials, across images, with Spearman-Brown correction (because models fitting used all trials per image). The model fit $r_\text{pred}$ was the correlation across test images between neuronal responses and model predictions. All experiments were conducted on a compute cluster with $300$ nodes, $48$ cores per node. CPU machines running Rocky Linux release 8.9 (Green Obsidian) were used. 

%% file: results.tex
\section{Results}\label{sec:results}
\begin{figure}[t]
\centering
        \includegraphics[width=\textwidth]{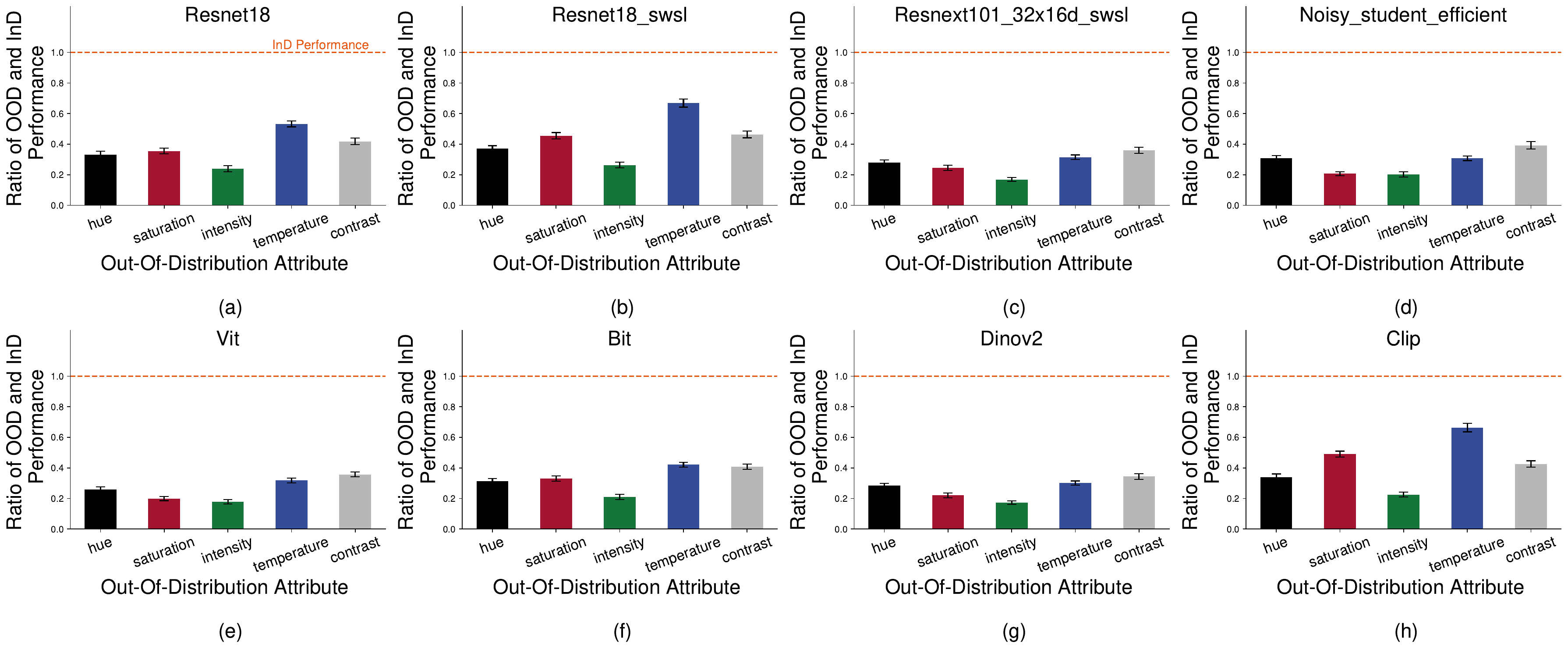}
        \caption{\textit{Neural predictivity drops under distribution shifts.} The y-axis shows the ratio of the neural predictivity for out-of-distribution (OOD) images to in-distribution (InD) test images. A ratio of $1$ would indicate no drop in performance. Each panel (a-h) shows a different architecture used for extracting image features. Each bar in a panels corresponds to a different OOD split constructed by using the \textit{high} hold-out strategy across $5$ different attributes (hue, saturation, saturation, intensity, temperature, and contrast). For all architectures and OOD splits, models fail to generalize well to OOD samples and are significantly and substantially below the $1.0$ horizontal line. Image features were extracted from the pre-final layer for all architectures.
        }
        \label{fig:ind_vs_ood}
\end{figure}

\begin{figure}[t]
\centering
        \includegraphics[width=\textwidth]{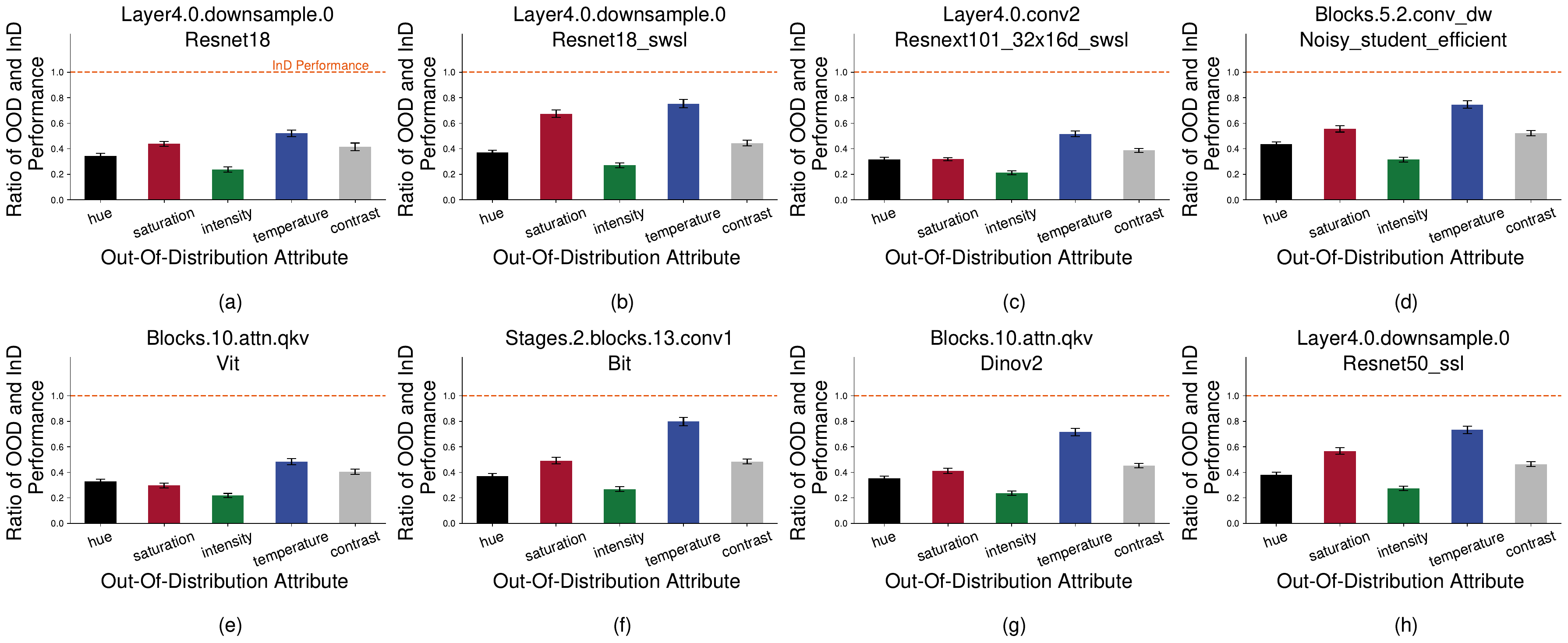}
        \caption{\textit{Neural predictivity drops for different model layers as well.} Neural predictivity on OOD samples is reported for multiple DNN architectures across multiple different layers. Layer name is mentioned alongside architecture in all panels (a-h). All OOD splits reported here were constructed using the \textit{high} hold-out strategy. For all architectures, layers, and OOD splits, models fail to generalize well to OOD samples and are significantly below the $1.0$ horizontal line.
        }
        \label{fig:ind_vs_ood_middle_layers}
\end{figure}

\begin{figure}[t]
\centering
        \includegraphics[width=\textwidth]{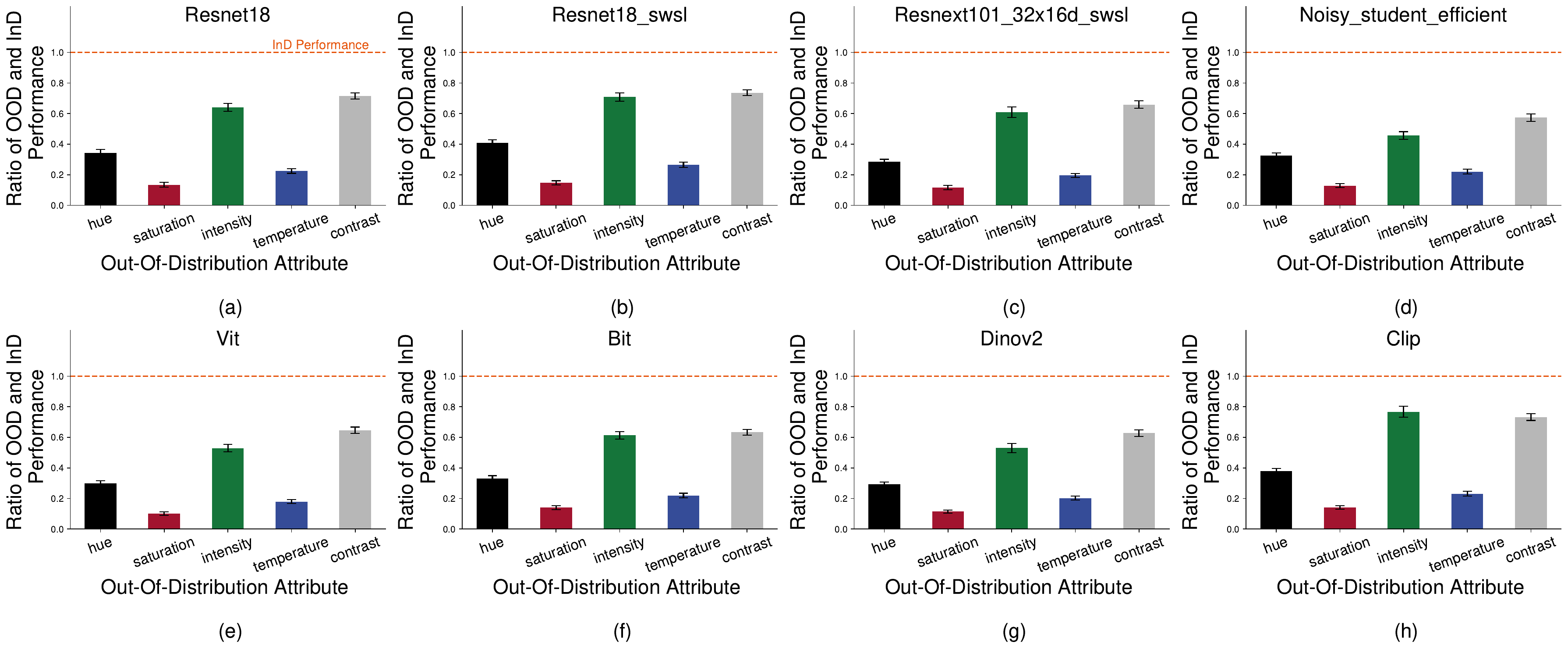}
        \caption{\textit{Neural predictivity drops for the low hold-out strategy as well.} Neural predictivity is reported on OOD test splits constructed using the \textit{low} hold-out strategy. Across all DNN architectures and image-computable attributes, performance is below 1.0 for all panels (a-h). Thus, models do not generalize well to OOD splits constructed with the \textit{low} hold-out strategy as well.
        }
        \label{fig:ind_vs_ood_high}
\end{figure}


\subsection{Neural prediticivity drops under distribution shifts}

DNN-based encoding models become worse at predicting neuronal responses under simple shifts in the image distribution. To demonstrate this, we report the ratio of neural predictivity between OOD and In-Distribution test splits ($r^2_{ood}/r^2_{ind}$). A ratio of 1 would indicate that models generalize equally well to InD and OOD test images (horizontal dashed line; Fig.~\ref{fig:ind_vs_ood}a). In contrast, the OOD/InD performance ratios are substantially lower than 1. For instance, the black bar in Fig.~\ref{fig:ind_vs_ood}a shows that the model's neural predictivity was $0.33$ on \textit{high}-hue OOD images (constructed using the \textit{high hold-out} strategy in Sec.~\ref{sec:constructing_ood}) compared to images with InD hue. Models show a similar lack of OOD generalization to OOD images with regard to saturation (red bar), intensity (green bar), temperature (blue bar), and contrast (gray bar). This performance drop was observed for all eight DNNs tested (Fig.~\ref{fig:ind_vs_ood}b-h) and ranged from a best-case ratio of $0.66$ for the CLIP model generalizing to \textit{high}-temperature OOD images to a worst-case ratio of $0.2$ for the ViT model generalizing to \textit{high}-saturation OOD images.

The lack of OOD generalization by neuron encoding models extended to models based on intermediate DNN layers, not just the penultimate layer. Fig.~\ref{fig:ind_vs_ood_middle_layers} reports OOD/InD generalization performance ratios of models based on activations extracted from intermediate DNN layers (layer names shown in Fig.~\ref{fig:ind_vs_ood_middle_layers}). For all architectures, OOD performance was substantially lower than InD performance. 

The lack of model OOD generalization extended to different hold-out strategies. Fig.~\ref{fig:ind_vs_ood_high} shows the OOD/InD model performance ratio for OOD splits constructed using the \textit{low} hold-out strategy described in Sec.~\ref{sec:constructing_ood}. OOD performance was lower than InD (ratios below $1$) for all architectures and image attributes. Additional results with the \textit{mid} hold-out strategy are provided in the supplement.

Combined, these results showcase a problem for current DNN-based models of the visual cortex---despite their ability to predict neural responses to in-distribution test images, the models generalize poorly under distribution shifts even in low-level image attributes. 

\subsection{The distance between train and test distributions explains generalization performance}
The results above raise a natural question---when and how do models of the ventral visual cortex fail to generalize under distribution shifts? Theoretical work has related OOD generalization to the amount of distribution shift ~\cite{canatar2021out,patil2024optimal}. Here we apply this theoretical framework to characterize generalization in DNN models of the brain. 
\begin{figure}[t]
\centering
        \includegraphics[width=\textwidth]{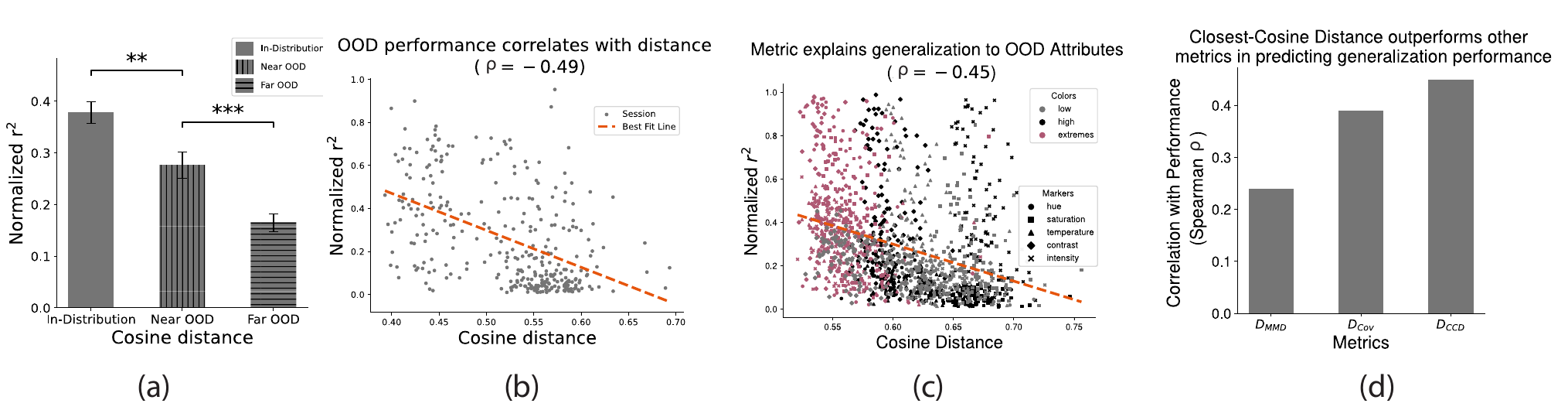}
        \caption{\textit{Closest-Cosine Distance metric well-explains performance across all attribute-based OOD splits.} (a) Neural predictivity on distance-based splits. Models performed best on In-Distribution (InD) the split, with a dip in performance from InD to Near OOD test set (two-sided t-test, $p<0.01$), and from Near OOD to Far-OOD (two-sided t-test, $p<0.01$). This suggests a relationship between the extent of distribution shift and generalization performance. (b) OOD performance can be well-explained by the distribution shift. For all $109$ sessions, the plot shows  performance on the InD, Near-OOD, and Far-OOD with the corresponding distribution shift measured using the Closest-Cosine Distance metric ($D_{CCD}$). Performance and $D_{CCD}$ have a Spearman correlation of $-0.49 (p<0.001)$. (c) Scatter plot of neural predictivity and the corresponding distribution shift ($D_{CCD}$) across all $15$ attribute-based OOD splits for all $109$ sessions. Generalization performance and the proposed distance metric have a Spearman correlation of $-0.45 (p<0.001)$ (d) Comparing different distance metrics w.r.t. their correlation with OOD performance. The proposed Closest-Cosine Distance has the highest correlation with neural predictivity, outperforming both MMD ($D_{MMD}$) and Covariate-Shift ($D_{Cov}$).
        }
        \label{fig:resnet_distance}
\end{figure}

Intuitively, model generalization should be worse for train-test splits under larger distribution shifts. We tested this intuition by constructing splits with different levels of distribution shifts---InD, Near OOD, and Far OOD. As described in Sec.~\ref{sec:constructing_ood}, images in every session were sorted based on cosine distance, and split into three chunks. The first chunk comprises the training and the In-Distribution test set, while the second and third chunks form the Near OOD and Far OOD test sets. As hypothesized, the model performance decreased progressively and significantly from In-Distribution to Near OOD, then Far OOD test distributions (Fig.~\ref{fig:resnet_distance}(a); two-sided t-test, $p<0.01$).

Beyond category-level differences, the size of the distribution shift predicted the OOD model performance drop across individual data splits  (Fig.~\ref{fig:resnet_distance}(b)). The distribution shift between each pair of train and OOD test distributions was quantified with the \textit{Closest Cosine Distance} ($D_{CCD}$; described in Sec.~\ref{sec:quantifying_ood}). The $D_{CCD}$ strongly correlated with the OOD model performance drop (Spearman correlation $\rho=-0.49$).

The distribution shift ($D_{CCD}$) calculated from ResNet features also explained OOD performance for attribute based splits (Fig.~\ref{fig:resnet_distance}(c). Across all image attributes (hue, saturation, temperature, contrast, intensity) and hold-out strategies (\textit{low, high, mid}) used to create OOD splits, $D_{CCD}$ correlated with OOD model performance drop with a Spearman correlation coefficient $\rho=-0.45$. Compared to two other popular measures of the sizes of distribution shifts (MMD, $D_{MMD}$\cite{gretton2012kernel} and Covariate-Shift, $D_{Cov}$~\cite{sugiyama2012machine}; Sec.~\ref{sec:quantifying_ood}), our proposed Closest Cosine Distance ($D_{CCD}$) metric best predicted OOD model performance (Fig.~\ref{fig:resnet_distance} (d)).

%% file: conclusions.tex
\section{Conclusions}\label{sec:conclusions}
These results reveal a deep problem in modern models of the visual cortex: good prediction is limited to the training image distribution. Simple distribution shifts break DNN models of the visual cortex, consistent with broader findings that the underlying DNNs are brittle to OOD shifts. Going one step further, we introduce an image-computable metric that significantly predicts the generalization performance of models under distribution shifts. This metric can help investigators gauge how well a neural model fit on one dataset may generalize to novel images. 

Our findings underline an important limitation of AI models for Neuroscience. Fields like Computer Vision have responded to the issue of distribution shifts by collecting progressive larger datasets, hoping models will learn to generalize to most images \cite{chen2024uncovering,caballero2022broken,prato2021scaling,naganuma2023empirical} at the billion-image scale. However, it is infeasible to achieve the same scale in neuroscience---the time needed to present a billion images is already a formidable challenge, not to mention the resource intensiveness of data collection. We hope our characterization of when and how modern models of the visual cortex fail out-of-domain will motivate the development of data-efficient ways to improve DNN generalization.

%% file: limitations.tex
\section{Limitations}\label{sec:limitations}
In this work, we have explored the impact of OOD samples on DNN-based models of the visual cortex. Our analyses have two main limitations that we hope future research can address. First, we did not fine-tune the DNNs on neural data. It is possible that training these models on the specific images and/or neural data can help improve generalization. Second, we did not explore the contributions of the images being OOD for the underlying pre-trained DNNs, as we only fit the linear encoding models on train set images and neural data. Because our images were naturalistic, it is plausible that they belonged to the training distribution of the pre-trained models we used, some of which (e.g., CLIP) having hundreds of millions of images. An interesting future direction will be to examine how the model performance is affected by using out-of-distribution images for the pre-trained DNNs. These images could include those from ImageNet-P, ImageNet-C~\cite{hendrycks2019benchmarking}, and evolved images~\cite{ponce2019evolving}.

%% file: supplement.tex
\section*{List of semantic categories in MacaqueITBench}
Table~\ref{table:semantic_categories} reports a list of all semantic categories in MacaqueITBench. As can be seen, the $8,233$ images correspond to over $300$ categories.
\input{category_table}
\section*{Sample Images from MacaqueITBench}
\begin{figure}[p]
\centering
        \includegraphics[width=\textwidth]{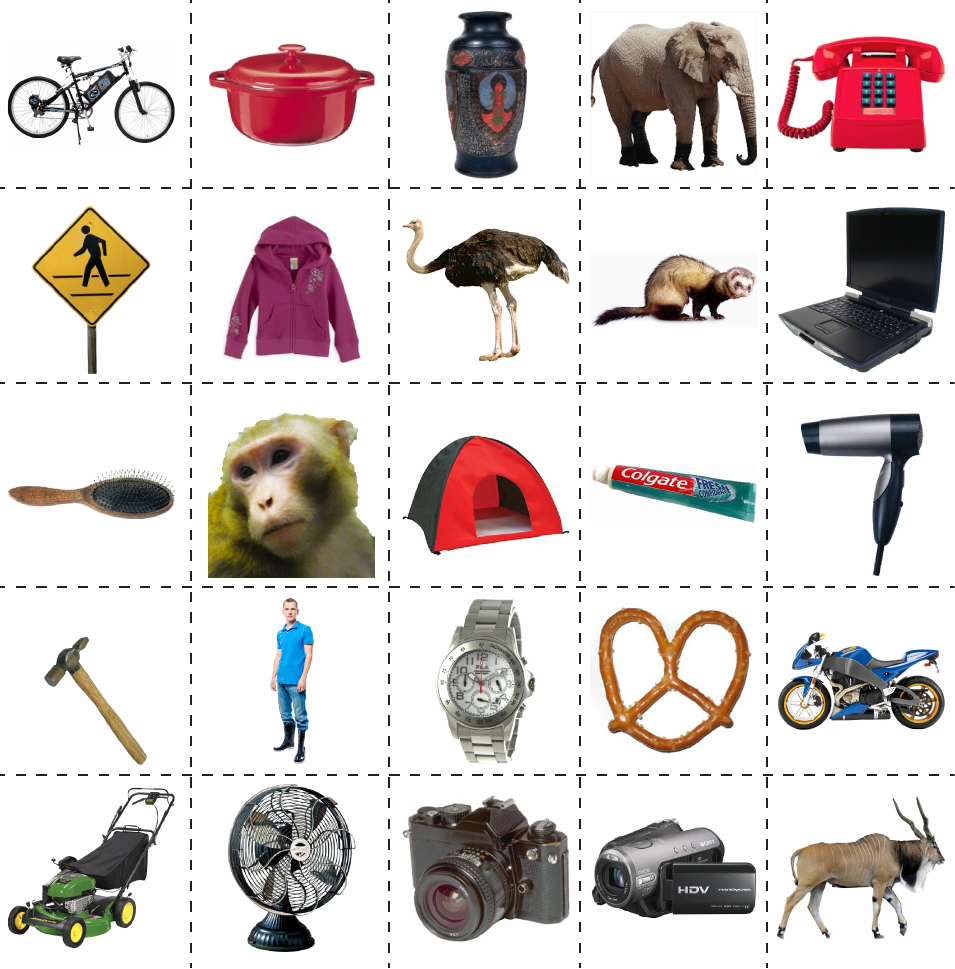}
        \caption{\textit{Images from MacaqueITBench.} 
        }
        \label{fig:sample_images}
\end{figure}

Fig.~\ref{fig:sample_images} shows sample images which were presented to Macaques to collect responses from the IT Cortex.
\section*{Additional results with Mid hold-out strategy}
\begin{figure}[p]
\centering
        \includegraphics[width=\textwidth]{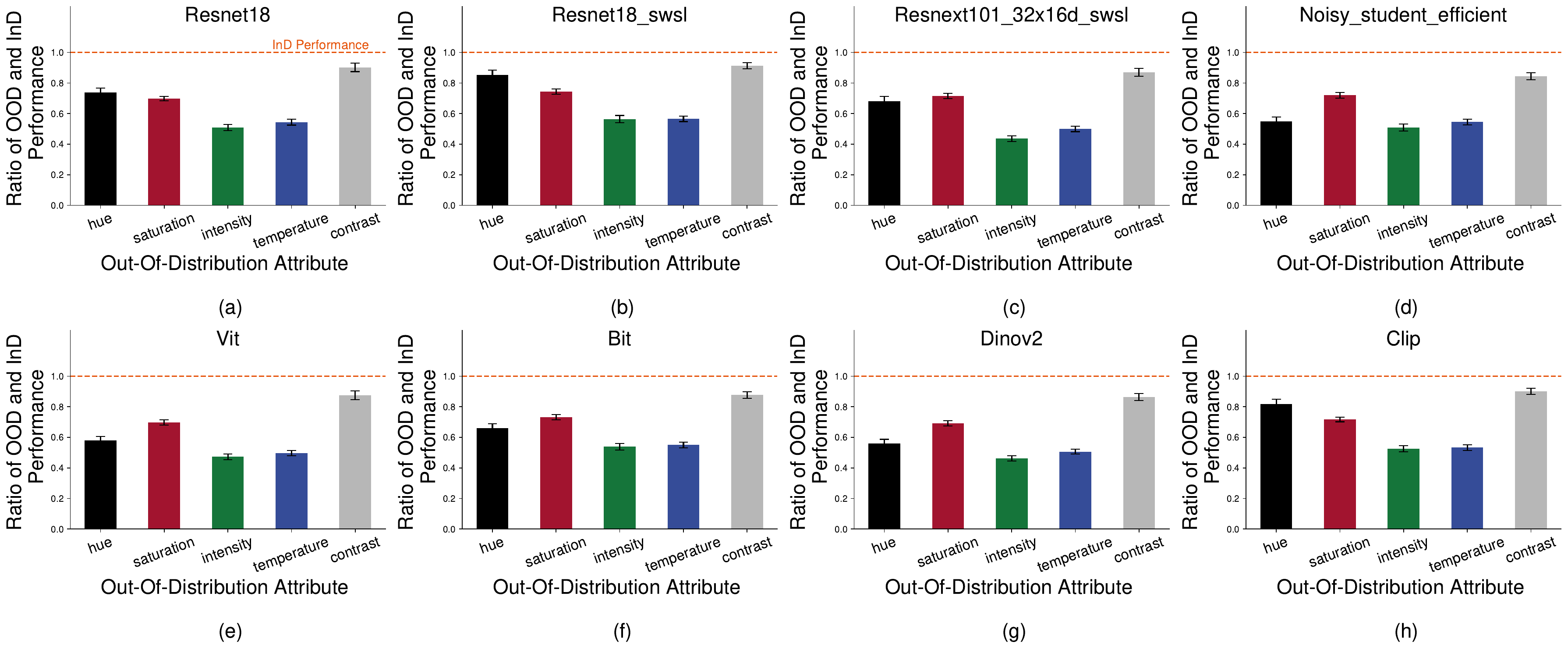}
        \caption{\textit{Neural predictivity drops for Mid hold-out strategy as well.} For all architectures, across multiple OOD shifts, performance on OOD is worse than in-distribution samples for the Mid hold-out strategy as well.
        }
        \label{fig:ind_vs_ood_mid}
\end{figure}

In the main paper, we presented results with two hold out strategies---high and low. Here, we present results with the third hold-out strategy outlined in the paper. We refer to this as the Mid hold out strategy as samples between the 42.5 and the 67.5 percentile of every OOD attribute are held out as the test set. As shown in Fig.~\ref{fig:ind_vs_ood_mid}, across all architectures and OOD attributes, models suffer to generalize to OOD samples for the Mid hold out strategy.

\section*{Additional results with intermediate layers}
\begin{figure}[p]
\centering
        \includegraphics[width=\textwidth]{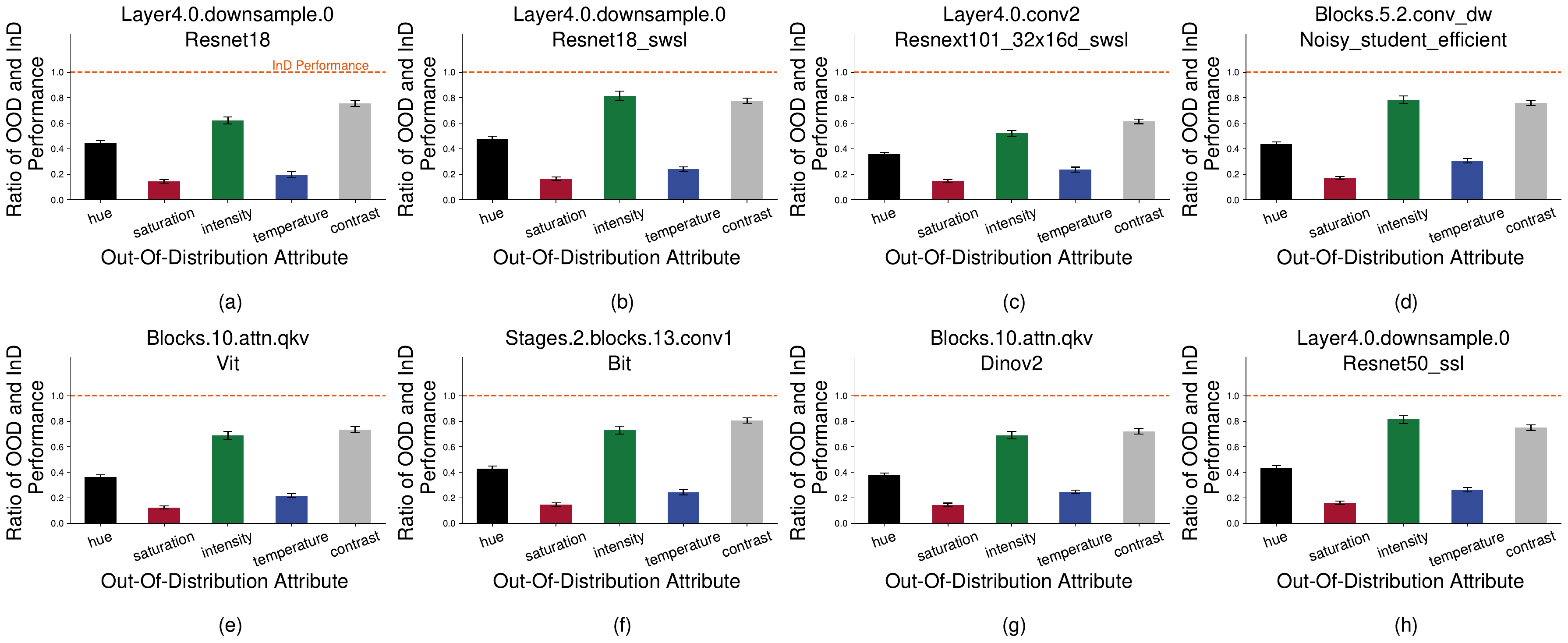}
        \caption{\textit{Neural predictivity drops for low hold-out strategy for intermediate layer features as well.} For all architectures, across multiple OOD shifts, performance on OOD is worse than in-distribution samples for the low hold-out strategy for image features extracted from intermediate DNN layers as well.
        }
        \label{fig:ind_vs_ood_low_middle}
\end{figure}

\begin{figure}[p]
\centering
        \includegraphics[width=\textwidth]{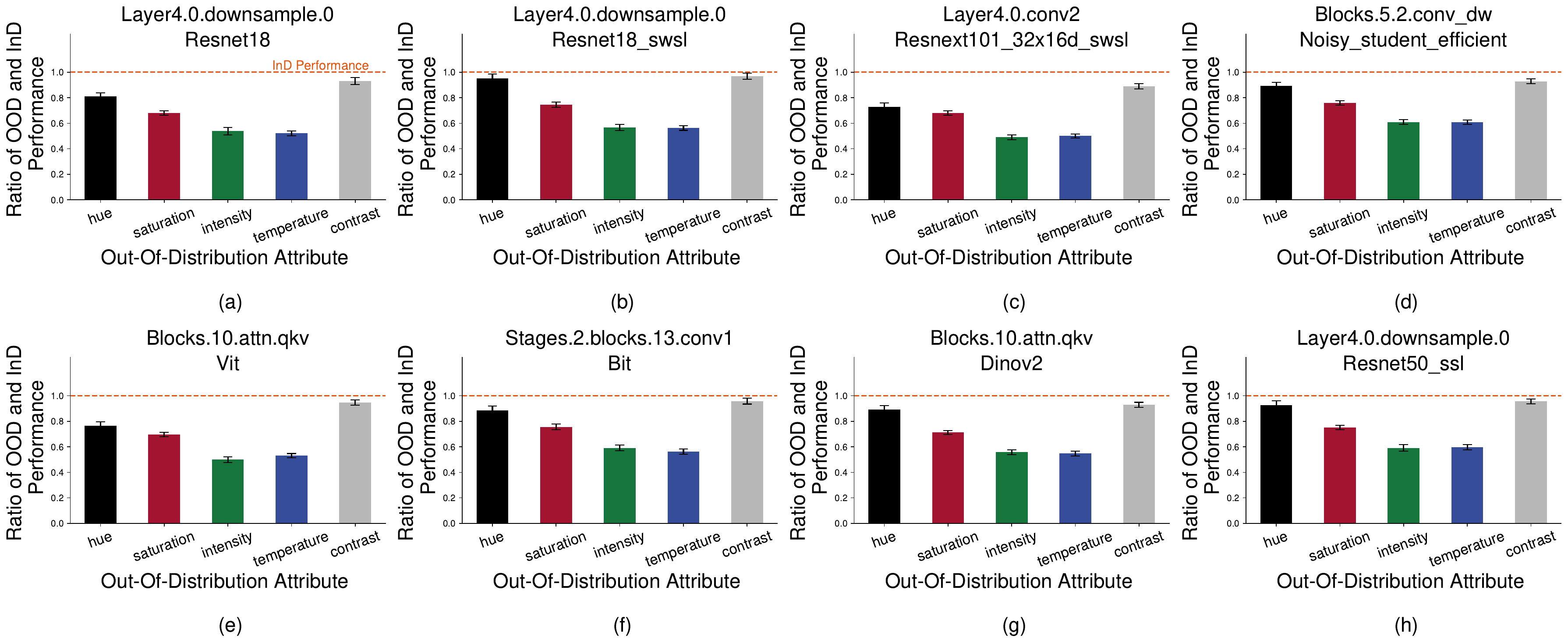}
        \caption{\textit{Neural predictivity drops for mid hold-out strategy for intermediate layer features as well.} For all architectures, across multiple OOD shifts, performance on OOD is worse than in-distribution samples for the mid hold-out strategy for image features extracted from intermediate DNN layers as well.
        }
        \label{fig:ind_vs_ood_mid_middle}
\end{figure}

In the main paper we presented results for models trained with intermediate layers for the high hold out strategy. Here we provide additional results with models that use intermediate layers of DNNs as feature extractors. In Fig.~\ref{fig:ind_vs_ood_low_middle} and Fig.~\ref{fig:ind_vs_ood_mid_middle} we report results for the \textit{low} and \textit{mid} hold-out strategies respectively.

%% file: category_table.tex
\begin{longtable}{|c|c|c|c|}
\hline
bottleopener & cupsaucer & tractor & bowtie \\
\hline
recordplayer & corkscrew & calculator & shoe \\
\hline
duster & lock & scissors & gift \\
\hline
guitar & swissarmyknife & pda & servingpiece \\
\hline
ceilingfan & nunchaku & radio & socks \\
\hline
sushi & filingcabinet & Other & waterbottle \\
\hline
leaves & watergun & trumpet & Big Animate \\
\hline
stove & chocolate & greenplant & bones \\
\hline
necktie & grapes & cookpot & windchime \\
\hline
cassettetape & mattress & fishbowl & Non \\
\hline
bench & tongs & microphone & lunchbox \\
\hline
jacket & bonzai & bullet & fan \\
\hline
cheesegrater & watch & cake & stool \\
\hline
pasta & sword & shirt & orientalplatesetting \\
\hline
typewriter & backpack & babushkadolls & hat \\
\hline
headphone & fork & wallsconce & hookah \\
\hline
boot & toothpaste & Gabor & abacus \\
\hline
quilt & short & familiarObjects & feather \\
\hline
fireplace & beermug & balloon & crossbow \\
\hline
pen & razor & dollhouse & carabiners \\
\hline
lightbulb & keychain & lawnmower & Glove \\
\hline
broom & headband & golfbag & garbagetrash \\
\hline
babyplayard & manorha & skateboard & shovel \\
\hline
christmasstocking & cooler & exercise & wineglassfull \\
\hline
camera & cheese & makeupcompact & plate \\
\hline
gong & cellphone & showercurtain & birdcage \\
\hline
tricycle & carfront & sleepingbag & window \\
\hline
umbrella & coatrack & roadsign & breadloaf \\
\hline
waxseal & mathcompass & dvdplayer & Rodent \\
\hline
handgun & binoculars & hilighter & icecreamcones \\
\hline
jack-o-lantern & basket & spoon & shredder \\
\hline
camcorder & christmastreeornamantball & apple & Face \\
\hline
log & cookingpan & scrunchie & stapler \\
\hline
flashlight & muffler & candy & orifan \\
\hline
golfball & pokercard & Bird & collar \\
\hline
washer & baseballcards & perfumebottle & babywalker \\
\hline
axe & patioloungechair & banana & wig \\
\hline
cookie & fish hook & motorcycle & sewingmachine \\
\hline
toy & pizza & lamp & meat \\
\hline
tape & tire & decorativescreen & musicstand \\
\hline
crib & candleholderwithcandle & grill & battery \\
\hline
hammer & compass & lei & hairdryer \\
\hline
giftbow & wheelbarrow & keyboard & trunk \\
\hline
iceskates & hanger & bathsuit & pill \\
\hline
Hand & kettle & microscope & Big \\
\hline
fruitparfait & Symbol & eraser & baseballbat \\
\hline
cage & lightswitch & laptop & sodacan \\
\hline
beaker & PPE & extra & kayak \\
\hline
sofa & fishingpole & microwave & mailbox \\
\hline
snowglobe & carseat & Butterfly & corset \\
\hline
doll & rollerskates & Fish & frisbee \\
\hline
trophy & saltpeppershake & pacifier & pezdispenser \\
\hline
rosary & router & airplane & Cat \\
\hline
reportfile & soapdispenser & coffin & yarn \\
\hline
dumbbell & chessboard & computer & aircompressor \\
\hline
birdhouse & Print & pipe & hotairballoon \\
\hline
doorknocker & anchor & bed & cashregister \\
\hline
loom & lipstick & measuringtape & chair \\
\hline
train & remotecontrol & toaster & coffeemug \\
\hline
pants & pie & donut & powerstrip \\
\hline
seasponge & beanbagchair & bike & domino \\
\hline
glasses & nailpolish & cherubstatue & knife \\
\hline
coin & printer & mp3player & leatherman \\
\hline
Turtle & flag & Toy & hairbrush \\
\hline
ladder & bucket & bell & ringbinder \\
\hline
wineglass & robot & stamp & spraybottle \\
\hline
mushroom & dresser & peppersonplate & tent \\
\hline
bowlofchips & videoGameController & lantern & candybar \\
\hline
cracker & computermouse & cane & ambulance \\
\hline
toothbrush & goggle & scooter & doorknob \\
\hline
gamesboard & lighter & tray & backgammon \\
\hline
tv & sink & doorwayarch & gamehandheld \\
\hline
wheelchair & objects & barbiedoll & coffeemaker \\
\hline
bagel & juice & shotglass & Mask \\
\hline
tablesmall & highchair & spoolofstring & helmet \\
\hline
horseshoe & telescope & hourglass & tweezer \\
\hline
ring & Misc & spicerack & handmirror \\
\hline
cushion & phone & vase & woodboxsmall \\
\hline
bowlingpin & clock & handbag & globe \\
\hline
key & muffin & dynamite & strainer \\
\hline
checkbook & pillow & sandwich & scale \\
\hline
ball & sippycup & Starfish & bottle \\
\hline
tupperware & cigarettepack & seashell & handheldvacuum \\
\hline
tree & earings & vacuum & suit \\
\hline
bullhorn & ketchupbottle & babycarriage & necklace \\
\hline
fridge & nest & slinky & curlingiron \\
\hline
desk & suitcase & pencilsharpener & speakers \\
\hline
button & rug & bowl & scroll \\
\hline
flask & paintbrush & bill & tennisracquet \\
\hline
boppypillow & rollingpin & saddle & frame \\
\hline
handkerchief & toiletseat & slate & licenseplate \\
\hline
laudrybasket & easteregg & accordian & crown \\
\hline
circuitboard & Dog & bongo & barrel \\
\hline
rock & pitcher & & \\
\hline
\caption{\textit{Images from MacaqueITBench.}}\label{table:semantic_categories}
\end{longtable}

%% file: neurips_2024.bbl
\begin{thebibliography}{10}

\bibitem{mcinnes2018umap}
Leland McInnes, John Healy, and James Melville.
\newblock Umap: Uniform manifold approximation and projection for dimension
  reduction.
\newblock {\em arXiv preprint arXiv:1802.03426}, 2018.

\bibitem{bashivan2019neural}
Pouya Bashivan, Kohitij Kar, and James~J DiCarlo.
\newblock Neural population control via deep image synthesis.
\newblock {\em Science}, 364(6439):eaav9436, 2019.

\bibitem{ponce2019evolving}
Carlos~R Ponce, Will Xiao, Peter~F Schade, Till~S Hartmann, Gabriel Kreiman,
  and Margaret~S Livingstone.
\newblock Evolving images for visual neurons using a deep generative network
  reveals coding principles and neuronal preferences.
\newblock {\em Cell}, 177(4):999--1009, 2019.

\bibitem{he2016deep}
Kaiming He, Xiangyu Zhang, Shaoqing Ren, and Jian Sun.
\newblock Deep residual learning for image recognition.
\newblock In {\em Proceedings of the IEEE conference on computer vision and
  pattern recognition}, pages 770--778, 2016.

\bibitem{yamins2014performance}
Daniel~LK Yamins, Ha~Hong, Charles~F Cadieu, Ethan~A Solomon, Darren Seibert,
  and James~J DiCarlo.
\newblock Performance-optimized hierarchical models predict neural responses in
  higher visual cortex.
\newblock {\em Proceedings of the national academy of sciences},
  111(23):8619--8624, 2014.

\bibitem{madan2023adversarial}
Spandan Madan, Tomotake Sasaki, Hanspeter Pfister, Tzu-Mao Li, and Xavier Boix.
\newblock Adversarial examples within the training distribution: A widespread
  challenge, 2023.

\bibitem{madan2022and}
Spandan Madan, Timothy Henry, Jamell Dozier, Helen Ho, Nishchal Bhandari,
  Tomotake Sasaki, Fr{\'e}do Durand, Hanspeter Pfister, and Xavier Boix.
\newblock When and how convolutional neural networks generalize to
  out-of-distribution category--viewpoint combinations.
\newblock {\em Nature Machine Intelligence}, 4(2):146--153, 2022.

\bibitem{cooper2021out}
Avi Cooper, Xavier Boix, Daniel Harari, Spandan Madan, Hanspeter Pfister,
  Tomotake Sasaki, and Pawan Sinha.
\newblock To which out-of-distribution object orientations are dnns capable of
  generalizing?
\newblock {\em arXiv preprint arXiv:2109.13445}, 2021.

\bibitem{madan2024improving}
Spandan Madan, You Li, Mengmi Zhang, Hanspeter Pfister, and Gabriel Kreiman.
\newblock Improving generalization by mimicking the human visual diet, 2024.

\bibitem{sakai2022three}
Akira Sakai, Taro Sunagawa, Spandan Madan, Kanata Suzuki, Takashi Katoh,
  Hiromichi Kobashi, Hanspeter Pfister, Pawan Sinha, Xavier Boix, and Tomotake
  Sasaki.
\newblock Three approaches to facilitate invariant neurons and generalization
  to out-of-distribution orientations and illuminations.
\newblock {\em Neural Networks}, 155:119--143, 2022.

\bibitem{hendrycks2019benchmarking}
Dan Hendrycks and Thomas Dietterich.
\newblock Benchmarking neural network robustness to common corruptions and
  perturbations.
\newblock {\em arXiv preprint arXiv:1903.12261}, 2019.

\bibitem{croce2023seasoning}
Francesco Croce, Sylvestre-Alvise Rebuffi, Evan Shelhamer, and Sven Gowal.
\newblock Seasoning model soups for robustness to adversarial and natural
  distribution shifts.
\newblock In {\em Proceedings of the IEEE/CVF Conference on Computer Vision and
  Pattern Recognition}, pages 12313--12323, 2023.

\bibitem{canatar2021out}
Abdulkadir Canatar, Blake Bordelon, and Cengiz Pehlevan.
\newblock Out-of-distribution generalization in kernel regression.
\newblock {\em Advances in Neural Information Processing Systems},
  34:12600--12612, 2021.

\bibitem{patil2024optimal}
Pratik Patil, Jin-Hong Du, and Ryan~J Tibshirani.
\newblock Optimal ridge regularization for out-of-distribution prediction.
\newblock {\em arXiv preprint arXiv:2404.01233}, 2024.

\bibitem{kriegeskorte2015deep}
Nikolaus Kriegeskorte.
\newblock Deep neural networks: a new framework for modeling biological vision
  and brain information processing.
\newblock {\em Annual review of vision science}, 1:417--446, 2015.

\bibitem{yamins2016using}
Daniel~LK Yamins and James~J DiCarlo.
\newblock Using goal-driven deep learning models to understand sensory cortex.
\newblock {\em Nature neuroscience}, 19(3):356--365, 2016.

\bibitem{schrimpf2018brain}
Martin Schrimpf, Jonas Kubilius, Ha~Hong, Najib~J Majaj, Rishi Rajalingham,
  Elias~B Issa, Kohitij Kar, Pouya Bashivan, Jonathan Prescott-Roy, Franziska
  Geiger, et~al.
\newblock Brain-score: Which artificial neural network for object recognition
  is most brain-like?
\newblock {\em BioRxiv}, page 407007, 2018.

\bibitem{ren2023well}
Yifei Ren and Pouya Bashivan.
\newblock How well do models of visual cortex generalize to out of distribution
  samples?
\newblock {\em bioRxiv}, pages 2023--05, 2023.

\bibitem{zhang2019making}
Richard Zhang.
\newblock Making convolutional networks shift-invariant again.
\newblock In {\em International conference on machine learning}, pages
  7324--7334. PMLR, 2019.

\bibitem{chaman2021truly}
Anadi Chaman and Ivan Dokmanic.
\newblock Truly shift-invariant convolutional neural networks.
\newblock In {\em Proceedings of the IEEE/CVF Conference on Computer Vision and
  Pattern Recognition}, pages 3773--3783, 2021.

\bibitem{mintun2021interaction}
Eric Mintun, Alexander Kirillov, and Saining Xie.
\newblock On interaction between augmentations and corruptions in natural
  corruption robustness.
\newblock {\em Advances in Neural Information Processing Systems}, 34, 2021.

\bibitem{xie2020adversarial}
Cihang Xie, Mingxing Tan, Boqing Gong, Jiang Wang, Alan~L Yuille, and Quoc~V
  Le.
\newblock Adversarial examples improve image recognition.
\newblock In {\em Proceedings of the IEEE/CVF Conference on Computer Vision and
  Pattern Recognition}, pages 819--828, 2020.

\bibitem{xie2020self}
Qizhe Xie, Minh-Thang Luong, Eduard Hovy, and Quoc~V Le.
\newblock Self-training with noisy student improves imagenet classification.
\newblock In {\em Proceedings of the IEEE/CVF conference on computer vision and
  pattern recognition}, pages 10687--10698, 2020.

\bibitem{madan2021small}
Spandan Madan, Tomotake Sasaki, Tzu-Mao Li, Xavier Boix, and Hanspeter Pfister.
\newblock Small in-distribution changes in 3d perspective and lighting fool
  both cnns and transformers.
\newblock {\em arXiv preprint arXiv:2106.16198}, 2021.

\bibitem{beery2018recognition}
Sara Beery, Grant Van~Horn, and Pietro Perona.
\newblock Recognition in terra incognita.
\newblock In {\em Proceedings of the European conference on computer vision
  (ECCV)}, pages 456--473, 2018.

\bibitem{zhang2021adversarial}
Qian Zhang, Qing Guo, Ruijun Gao, Felix Juefei-Xu, Hongkai Yu, and Wei Feng.
\newblock Adversarial relighting against face recognition.
\newblock {\em arXiv preprint arXiv:2108.07920}, 2021.

\bibitem{barbu2019objectnet}
Andrei Barbu, David Mayo, Julian Alverio, William Luo, Christopher Wang, Dan
  Gutfreund, Josh Tenenbaum, and Boris Katz.
\newblock Objectnet: A large-scale bias-controlled dataset for pushing the
  limits of object recognition models.
\newblock {\em Advances in neural information processing systems}, 32, 2019.

\bibitem{liu2018beyond}
Hsueh-Ti~Derek Liu, Michael Tao, Chun-Liang Li, Derek Nowrouzezahrai, and Alec
  Jacobson.
\newblock Beyond pixel norm-balls: Parametric adversaries using an analytically
  differentiable renderer.
\newblock {\em arXiv preprint arXiv:1808.02651}, 2018.

\bibitem{zeng2019adversarial}
Xiaohui Zeng, Chenxi Liu, Yu-Siang Wang, Weichao Qiu, Lingxi Xie, Yu-Wing Tai,
  Chi-Keung Tang, and Alan~L Yuille.
\newblock Adversarial attacks beyond the image space.
\newblock In {\em Proceedings of the IEEE/CVF Conference on Computer Vision and
  Pattern Recognition}, pages 4302--4311, 2019.

\bibitem{sakai2021three}
Akira Sakai, Taro Sunagawa, Spandan Madan, Kanata Suzuki, Takashi Katoh,
  Hiromichi Kobashi, Hanspeter Pfister, Pawan Sinha, Xavier Boix, and Tomotake
  Sasaki.
\newblock Three approaches to facilitate dnn generalization to objects in
  out-of-distribution orientations and illuminations: late-stopping, tuning
  batch normalization and invariance loss.
\newblock {\em arXiv preprint arXiv:2111.00131}, 2021.

\bibitem{belder2022random}
Amir Belder, Gal Yefet, Ran Ben-Itzhak, and Ayellet Tal.
\newblock Random walks for adversarial meshes.
\newblock In {\em ACM SIGGRAPH 2022 Conference Proceedings}, pages 1--9, 2022.

\bibitem{xiao2019meshadv}
Chaowei Xiao, Dawei Yang, Bo~Li, Jia Deng, and Mingyan Liu.
\newblock Meshadv: Adversarial meshes for visual recognition.
\newblock In {\em Proceedings of the IEEE/CVF Conference on Computer Vision and
  Pattern Recognition}, pages 6898--6907, 2019.

\bibitem{yang2018realistic}
Dawei Yang, Chaowei Xiao, Bo~Li, Jia Deng, and Mingyan Liu.
\newblock Realistic adversarial examples in 3d meshes.
\newblock {\em arXiv preprint arXiv:1810.05206}, 2:2, 2018.

\bibitem{joshi2019semantic}
Ameya Joshi, Amitangshu Mukherjee, Soumik Sarkar, and Chinmay Hegde.
\newblock Semantic adversarial attacks: Parametric transformations that fool
  deep classifiers.
\newblock In {\em Proceedings of the IEEE/CVF international conference on
  computer vision}, pages 4773--4783, 2019.

\bibitem{shamsabadi2020colorfool}
Ali~Shahin Shamsabadi, Ricardo Sanchez-Matilla, and Andrea Cavallaro.
\newblock Colorfool: Semantic adversarial colorization.
\newblock In {\em Proceedings of the IEEE/CVF Conference on Computer Vision and
  Pattern Recognition}, pages 1151--1160, 2020.

\bibitem{bomatter2021pigs}
Philipp Bomatter, Mengmi Zhang, Dimitar Karev, Spandan Madan, Claire Tseng, and
  Gabriel Kreiman.
\newblock When pigs fly: Contextual reasoning in synthetic and natural scenes.
\newblock {\em arXiv preprint arXiv:2104.02215}, 2021.

\bibitem{zhang2020putting}
Mengmi Zhang, Claire Tseng, and Gabriel Kreiman.
\newblock Putting visual object recognition in context.
\newblock In {\em Proceedings of the IEEE/CVF Conference on Computer Vision and
  Pattern Recognition}, pages 12985--12994, 2020.

\bibitem{arjovsky2019invariant}
Martin Arjovsky, L{\'e}on Bottou, Ishaan Gulrajani, and David Lopez-Paz.
\newblock Invariant risk minimization.
\newblock {\em arXiv preprint arXiv:1907.02893}, 2019.

\bibitem{erfani2016robust}
Sarah Erfani, Mahsa Baktashmotlagh, Masud Moshtaghi, Xuan Nguyen, Christopher
  Leckie, James Bailey, and Rao Kotagiri.
\newblock Robust domain generalisation by enforcing distribution invariance.
\newblock In {\em Proceedings of the Twenty-Fifth International Joint
  Conference on Artificial Intelligence (IJCAI-16)}, pages 1455--1461. AAAI
  Press, 2016.

\bibitem{chattopadhyay2020learning}
Prithvijit Chattopadhyay, Yogesh Balaji, and Judy Hoffman.
\newblock Learning to balance specificity and invariance for in and out of
  domain generalization.
\newblock In {\em Computer Vision--ECCV 2020: 16th European Conference,
  Glasgow, UK, August 23--28, 2020, Proceedings, Part IX 16}, pages 301--318.
  Springer, 2020.

\bibitem{wang2021respecting}
Ziqi Wang, Marco Loog, and Jan Van~Gemert.
\newblock Respecting domain relations: Hypothesis invariance for domain
  generalization.
\newblock In {\em 2020 25th International Conference on Pattern Recognition
  (ICPR)}, pages 9756--9763. IEEE, 2021.

\bibitem{li2017deeper}
Da~Li, Yongxin Yang, Yi-Zhe Song, and Timothy~M Hospedales.
\newblock Deeper, broader and artier domain generalization.
\newblock In {\em Proceedings of the IEEE international conference on computer
  vision}, pages 5542--5550, 2017.

\bibitem{li2018learning}
Da~Li, Yongxin Yang, Yi-Zhe Song, and Timothy Hospedales.
\newblock Learning to generalize: Meta-learning for domain generalization.
\newblock In {\em Proceedings of the AAAI conference on artificial
  intelligence}, volume~32, 2018.

\bibitem{balaji2018metareg}
Yogesh Balaji, Swami Sankaranarayanan, and Rama Chellappa.
\newblock Metareg: Towards domain generalization using meta-regularization.
\newblock {\em Advances in neural information processing systems}, 31, 2018.

\bibitem{zhou2021domain}
Kaiyang Zhou, Ziwei Liu, Yu~Qiao, Tao Xiang, and Chen~Change Loy.
\newblock Domain generalization in vision: A survey.
\newblock {\em arXiv preprint arXiv:2103.02503}, 2021.

\bibitem{zhang2017mixup}
Hongyi Zhang, Moustapha Cisse, Yann~N Dauphin, and David Lopez-Paz.
\newblock mixup: Beyond empirical risk minimization.
\newblock {\em arXiv preprint arXiv:1710.09412}, 2017.

\bibitem{hendrycks2019augmix}
Dan Hendrycks, Norman Mu, Ekin~D Cubuk, Barret Zoph, Justin Gilmer, and Balaji
  Lakshminarayanan.
\newblock Augmix: A simple data processing method to improve robustness and
  uncertainty.
\newblock {\em arXiv preprint arXiv:1912.02781}, 2019.

\bibitem{xu2020randconv}
Zhenlin Xu, Deyi Liu, Junlin Yang, Colin Raffel, and Marc Niethammer.
\newblock Robust and generalizable visual representation learning via random
  convolutions.
\newblock {\em arXiv preprint arXiv:2007.13003}, 2020.

\bibitem{huang2017adain}
Xun Huang and Serge Belongie.
\newblock Arbitrary style transfer in real-time with adaptive instance
  normalization.
\newblock In {\em Proceedings of the IEEE international conference on computer
  vision}, pages 1501--1510, 2017.

\bibitem{shankar2018generalizing}
Shiv Shankar, Vihari Piratla, Soumen Chakrabarti, Siddhartha Chaudhuri, Preethi
  Jyothi, and Sunita Sarawagi.
\newblock Generalizing across domains via cross-gradient training.
\newblock {\em arXiv preprint arXiv:1804.10745}, 2018.

\bibitem{qiao2020learning}
Fengchun Qiao, Long Zhao, and Xi~Peng.
\newblock Learning to learn single domain generalization.
\newblock In {\em Proceedings of the IEEE/CVF Conference on Computer Vision and
  Pattern Recognition}, pages 12556--12565, 2020.

\bibitem{sinha2017certifying}
Aman Sinha, Hongseok Namkoong, Riccardo Volpi, and John Duchi.
\newblock Certifying some distributional robustness with principled adversarial
  training.
\newblock {\em arXiv preprint arXiv:1710.10571}, 2017.

\bibitem{NEURIPS2018_1d94108e}
Riccardo Volpi, Hongseok Namkoong, Ozan Sener, John~C Duchi, Vittorio Murino,
  and Silvio Savarese.
\newblock Generalizing to unseen domains via adversarial data augmentation.
\newblock In S.~Bengio, H.~Wallach, H.~Larochelle, K.~Grauman, N.~Cesa-Bianchi,
  and R.~Garnett, editors, {\em Advances in Neural Information Processing
  Systems}, volume~31. Curran Associates, Inc., 2018.

\bibitem{billion_scale}
I.~Zeki Yalniz, Herv{\'{e}} J{\'{e}}gou, Kan Chen, Manohar Paluri, and Dhruv
  Mahajan.
\newblock Billion-scale semi-supervised learning for image classification.
\newblock {\em CoRR}, abs/1905.00546, 2019.

\bibitem{radford2021learning}
Alec Radford, Jong~Wook Kim, Chris Hallacy, Aditya Ramesh, Gabriel Goh,
  Sandhini Agarwal, Girish Sastry, Amanda Askell, Pamela Mishkin, Jack Clark,
  et~al.
\newblock Learning transferable visual models from natural language
  supervision.
\newblock In {\em International conference on machine learning}, pages
  8748--8763. PMLR, 2021.

\bibitem{oquab2023dinov2}
Maxime Oquab, Timoth{\'e}e Darcet, Th{\'e}o Moutakanni, Huy Vo, Marc
  Szafraniec, Vasil Khalidov, Pierre Fernandez, Daniel Haziza, Francisco Massa,
  Alaaeldin El-Nouby, et~al.
\newblock Dinov2: Learning robust visual features without supervision.
\newblock {\em arXiv preprint arXiv:2304.07193}, 2023.

\bibitem{konkle2010conceptual}
Talia Konkle, Timothy~F Brady, George~A Alvarez, and Aude Oliva.
\newblock Conceptual distinctiveness supports detailed visual long-term memory
  for real-world objects.
\newblock {\em Journal of experimental Psychology: general}, 139(3):558, 2010.

\bibitem{buzsaki2012origin}
Gy{\"o}rgy Buzs{\'a}ki, Costas~A Anastassiou, and Christof Koch.
\newblock The origin of extracellular fields and currents—eeg, ecog, lfp and
  spikes.
\newblock {\em Nature reviews neuroscience}, 13(6):407--420, 2012.

\bibitem{super2005chronic}
Hans Super and Pieter~R Roelfsema.
\newblock Chronic multiunit recordings in behaving animals: advantages and
  limitations.
\newblock {\em Progress in brain research}, 147:263--282, 2005.

\bibitem{dalal2005histograms}
Navneet Dalal and Bill Triggs.
\newblock Histograms of oriented gradients for human detection.
\newblock In {\em 2005 IEEE computer society conference on computer vision and
  pattern recognition (CVPR'05)}, volume~1, pages 886--893. Ieee, 2005.

\bibitem{oliva2006building}
Aude Oliva and Antonio Torralba.
\newblock Building the gist of a scene: The role of global image features in
  recognition.
\newblock {\em Progress in brain research}, 155:23--36, 2006.

\bibitem{dosovitskiy2020image}
Alexey Dosovitskiy, Lucas Beyer, Alexander Kolesnikov, Dirk Weissenborn,
  Xiaohua Zhai, Thomas Unterthiner, Mostafa Dehghani, Matthias Minderer, Georg
  Heigold, Sylvain Gelly, et~al.
\newblock An image is worth 16x16 words: Transformers for image recognition at
  scale.
\newblock {\em arXiv preprint arXiv:2010.11929}, 2020.

\bibitem{xie2020selftraining}
Qizhe Xie, Minh-Thang Luong, Eduard Hovy, and Quoc~V. Le.
\newblock Self-training with noisy student improves imagenet classification,
  2020.

\bibitem{xiao2024feature}
Will Xiao, Saloni Sharma, Gabriel Kreiman, and Margaret~S Livingstone.
\newblock Feature-selective responses in macaque visual cortex follow eye
  movements during natural vision.
\newblock {\em Nature Neuroscience}, pages 1--10, 2024.

\bibitem{gretton2012kernel}
Arthur Gretton, Karsten~M Borgwardt, Malte~J Rasch, Bernhard Sch{\"o}lkopf, and
  Alexander Smola.
\newblock A kernel two-sample test.
\newblock {\em The Journal of Machine Learning Research}, 13(1):723--773, 2012.

\bibitem{sugiyama2012machine}
Masashi Sugiyama and Motoaki Kawanabe.
\newblock {\em Machine learning in non-stationary environments: Introduction to
  covariate shift adaptation}.
\newblock MIT press, 2012.

\bibitem{chen2024uncovering}
Dingshuo Chen, Yanqiao Zhu, Jieyu Zhang, Yuanqi Du, Zhixun Li, Qiang Liu, Shu
  Wu, and Liang Wang.
\newblock Uncovering neural scaling laws in molecular representation learning.
\newblock {\em Advances in Neural Information Processing Systems}, 36, 2024.

\bibitem{caballero2022broken}
Ethan Caballero, Kshitij Gupta, Irina Rish, and David Krueger.
\newblock Broken neural scaling laws.
\newblock {\em arXiv preprint arXiv:2210.14891}, 2022.

\bibitem{prato2021scaling}
Gabriele Prato, Simon Guiroy, Ethan Caballero, Irina Rish, and Sarath Chandar.
\newblock Scaling laws for the out-of-distribution generalization of image
  classifiers.
\newblock In {\em ICML 2021 Workshop on Uncertainty and Robustness in Deep
  Learning}, 2021.

\bibitem{naganuma2023empirical}
Hiroki Naganuma and Ryuichiro Hataya.
\newblock An empirical investigation of pre-trained model selection for
  out-of-distribution generalization and calibration.
\newblock {\em arXiv preprint arXiv:2307.08187}, 2023.

\end{thebibliography}
